\documentstyle[12pt]{article}
\begin{document}
\newcommand{\w}{\wedge}
\newcommand{\sh}{S^{\w}}
\newcommand{\two}{(\hskip-5pt\begin{array}{c}{\scriptstyle
2}\\[-8pt]{\scriptstyle\cdot}\end{array}\hskip-5pt)}
\newcommand{\sqc}{S^2_{qc}}
\newcommand{\su}{SU_q(2)}
\newcommand{\ap}{{\cal A}}
\newcommand{\th}{\theta}
\newcommand{\ce}{{\bf C}}
\newcommand{\nn}{{\bf N}}
\newcommand{\lo}{\longrightarrow}
\newcommand{\p}{\mathop\otimes}
\newcommand{\lon}{\Longleftrightarrow}
\newcommand{\om}{\omega}
\newcommand{\sm}{\mathop\oplus}
\setbox1=\hbox{$\top$}
\newcommand{\tp}{\smash{\bigcirc}\hskip-10.6pt\raise-2pt\copy1\,}
\newcommand{\ti}{\smash{\bigcirc}\!\!\!\!\!\!\!\perp}
\newcommand{\tiq}{\smash{\bigcirc}\!\!\!\!\!\!\!\perp}
\newcommand{\la}{\lambda}
\newcommand{\spr}{{\scriptstyle \circ}}
\newcommand{\noi}{\noindent}
\newcommand{\al}{\alpha}
\newcommand{\nm}{\parallel}
\newcommand{\nsubset}{\not\subset}
\title{Symmetries of quantum spaces. Subgroups and quotient
spaces of quantum $SU(2)$ and $SO(3)$ groups.}
\author{Piotr Podle\'s\\
Department of Mathematical Methods in Physics,\\ Faculty of
Physics, Warsaw University,\\ Ho\.za 74,
00-682 Warszawa,
Poland}
\maketitle
\begin{abstract}
We prove that each action of a compact matrix
quantum group on a compact quantum
space can be decomposed into irreducible representations of the group. We
give the formula for the corresponding multiplicities in the case of
the quotient quantum spaces. We describe the subgroups and the quotient
spaces of quantum SU(2) and SO(3) groups.
\end{abstract}
\addtocounter{section}{-1}
\section{Introduction}

\ \ \ \ Quantum groups have been already applied in various areas of physics,
 like conformal field theory and exactly solvable models in statistical
mechanics. It is especially interesting that they could
possibly describe symmetries of (quantum) space-time in a future quantum
gravity. In the same time, the nature and properties of quantum groups are
still under investigation. The local description of quantum groups is
given in terms of quantum universal enveloping algebras (cf e.g. [Dr],
[J]). In the global description we investigate the functions on quantum
groups (cf e.g. [W2], [RTF]). A deep insight in that global structure is
given by the topological approach developed in the series of papers of
S.L. Woronowicz [W1]-[W6]. We use that approach in the present paper.

The classical $SU(2)$ and $SO(3)$ groups play an important role in
description of spherically symmetric, stationary problems in physics. Also
their subgroups are important in description of various physical systems.
The description of quantum $SU(2)$ groups was given in [W2]. Their quantum
homogeneous spaces, quantum 2-spheres, were investigated in [P1], [P2], [P5]
(cf also [VS2]). However,
the general theory of quantum subgroups and quantum homogeneous spaces was
only touched there. In the present paper we want to treat that subject im
more detail. We also provide more examples.

In Section 1 we investigate the general theory of the (right) actions
of (compact matrix) quantum groups on (compact) quantum spaces. In
Sections 2 and 3 the theory is illustrated on the example of quantum
$SU(2)$ and $SO(3)$ groups. We classify their subgroups and describe the
corresponding quotient spaces. Provided examples of finite quantum groups
can have an application in the theory of pseudogroups of Ocneanu.
In the course of the paper we substanciate
some statements made in [P1] and [P5]. The results of the paper were
partially contained in [P3] and partially announced in [P4].

Throughout
the paper we use the terminology and results of [W2], [W3]. All considered
$C^*$-algebras and $C^*$-homomorphisms are unital.
The symbol $\approx$ denotes a $C^*$-isomorphism.
If $M$ is a subset of a
$C^*$-algebra $A$ then $<M>$ denotes the closure of $\ \mbox{span}\ M$.
Let us recall (cf [W1]) that (compact) quantum spaces $X$ are abstract
objects which are in bijective correspondence with $C^*$-algebras $C(X)$.
In particular, if $X$ is a usual (compact
Hausdorff) space then $C(X)$ has the usual
meaning of $C^*$-algebra of continuous functions on $X$. Each commutative
$C^*$-algebra can be obtained in that way (up to a $C^*$-isomorphism).

We
use the Pauli matrices
\[ \sigma_x=\left[\begin{array}{ll}
0&1\\1&0
\end{array}\right],\
\sigma_y=\left[\begin{array}{ll}
0&-i\\i&0
\end{array}\right],\
\sigma_z=\left[\begin{array}{ll}
1&0\\0&-1
\end{array}\right]. \]
We sum over repeated indices which are not taken in brackets (Einstein's
convention). For $x\in{\bf R}$, $E(x)$ denotes the integer part of $x$.\\\\

\section{Symmetries of quantum spaces}

In this Section
we define the notion of subgroup of (compact matrix) quantum
group. We also provide the basic notions concerning the actions of quantum
groups on (compact)
quantum spaces. We prove that each such action can be decomposed into
irreducible representations of the quantum group. We give the formula for
the corresponding multiplicities in the case of quotient quantum spaces.

Let us recall\\\\
{\sc Definition 1.1} ([W3], [W7])

\begin{em}
$G=(A,u)$ is called a (compact matrix) quantum group if $A\neq\{0\}$
is a $C^*$-algebra, $u=(u_{ij})_{i,j=1}^N$
 is an $N\times N$ matrix with entries
belonging to $A$ and
\begin{enumerate}
\item $A$ is the smallest $C^*$-algebra containing all
matrix elements of $u$
\item there exists a $C^*$-algebra homomorphism $\Phi:A\lo A\p
A$ such that
\begin{equation}
\Phi(u_{kl})=\sum^N_{r=1} u_{kr}\p u_{rl}\qquad k,l=1,2,\ldots,N, \label{1.1}
\end{equation}
\item $u$ and $u^T=(u_{lk})^N_{k,l=1}$ are invertible.\\\\
\end{enumerate}
\end{em}

In particular, each compact group of matrices $G\subset GL(N,\ce)$ is a
quantum group [W3]. Then $A=C(G)$ and $u$ corresponds to the fundamental
representation of $G$: $u_{ij}(g)=g_{ij}\in\ce$, $g\in G$, $i,j=1,\ldots,N$.
Each quantum group with commutative $A$ is of that kind (up to a
$C^*$-isomorphism). We use the notation $A=C(G)$ for any quantum group.
We say [W3] that $w$ is a (smooth nondegenerate) representation of $G$ if $w$
is an invertible $M\times M$ matrix with entries in $A$ and
\[ \Phi(w_{kl})=\sum^N_{r=1} w_{kr}\p w_{rl},\qquad k,l=1,2,\ldots,M, \]
for some $M\in\nn$. We denote $M=\dim\ w$.
It is easy to see (cf [W3]) that $w^T$ is also invertible
and therefore the $w$-image of $G$ is a quantum group:\\\\
{\sc Proposition 1.2}

\begin{em}
Let $w$ be a representation of a quantum group, $M=\dim\ w$. Then
\[ (C^*(\{w_{ij}: i,j=1,\ldots,M\}), w) \]
is also a quantum group.\\\\
\end{em}
{\sc Note.} Let $C^*(\{w_{ij}: i,j=1,\ldots,M\})=A$. Then quantum
groups $(A,u)$ and $(A,w)$ have the same $\Phi$ and can be identified.\\

The unital $*$-algebra  generated by all matrix elements
of $u$ is denoted by $\ap$.
Tensor product ($\tp$), direct sum ($\sm$), equivalence ($\simeq$) and
irreducibility of representations of $G$ are defined as for
usual matrices (cf [W3]). In particular, representations $w$, $w'$
are equivalent if $\dim w=\dim w'$ and there exists $S\in
GL(\dim w,\ce)$ such that $w=Sw'S^{-1}$.
Each representation is equivalent to a representation which is unitary
(as matrix).
Let $\{u^{\tau}\}_{\tau\in\hat G}$ be the set of all nonequivalent
irreducible unitary representations of $G$. We denote by $u^0$ the trivial
representation ($0\in\hat G$,
$\dim u^0=1$ and $u^0_{11}=I$). Set $d_{\tau}=\dim u^{\tau}$.
Due to [W3, Prop.4.7],
the matrix elements of all $u^{\tau}$, $\tau\in\hat G$ give a linear
basis
of $\ap$.
The Haar measure $h$ is the state on $C(G)$ which is equal $1$ on $I$
and 0 on other matrix elements of
$u^{\tau}$, $\tau\in\hat G$.
It is invariant, i.e. $(id\p h)\Phi(x)=(h\p id)\Phi(x)=h(x)I$,
$x\in C(G)$ [W3, Th.4.2]. According to (5.10) and (5.15) of [W3],
there exist matrices $F_{\alpha}$, $\alpha\in\hat G$,
such that
$$h({u^{\alpha}_{km}}^*u^{\beta}_{ln})=(Tr F_{(\alpha)})^{-1}%
\delta_{\alpha\beta}\delta_{mn} (F_{(\alpha)}^{-1})_{lk}\ \ \ .$$
We set
$x^{\alpha}_{sm}=(Tr F_{(\alpha)}){u^{\alpha}_{km}}^*(F_{(\alpha)})_{ks}$,
$\rho^{\alpha}_{sm}(x)=h(x^{\alpha}_{sm}x)$ for $x\in C(G)$,
$\alpha\in\hat G$, $s,m=1,\ldots,d_{\alpha}$. Then $\rho^{\alpha}_{sm}$
are continuous linear functionals on $C(G)$ and
$\rho^{\alpha}_{sm}(u^{\beta}_{ln})=%
\delta_{\alpha\beta}\delta_{sl}\delta_{mn}$,\ \ \  $\beta\in\hat G$,
$l,n=1,2,\ldots,d_{\beta}$. Hence,
\begin{equation}
(\rho^{\alpha}_{sm}\p\rho^{\beta}_{ij})\Phi_G=%
\delta_{mi}\delta_{\alpha(\beta)}\rho^{\beta}_{sj}\label{1.2}
\end{equation}
(the action of both sides on all $u^{\tau}_{ab}$ is the same). We put
$P^{\alpha}_{sm}=(id\p\rho^{\alpha}_{sm})\Phi_G$,
$\rho^{\alpha}=\rho^{\alpha}_{ss}\in C(G)'$ (Einstein's convention!). Then
\begin{equation}
P^{\alpha}_{sm}C(G)\subset\ \mbox{span}\ \{u^{\alpha}_{is}: i=1,2,\ldots,%
d_{\alpha}\},\label{1.3}
\end{equation}
\begin{equation}
\rho^{\alpha}(u^{\beta}_{ln})=\delta_{\alpha\beta}\delta_{ln}\label{1.4}
\end{equation}
(cf [W3, eq.5.37]). In particular $\rho^0=h$. The basic notion of this Section
is given by\\\\

{\sc Definition 1.3}

\begin{em}
We say that a quantum group $H=(B,v)$ is a (compact)
subgroup of a quantum group
$G=(A,u)$ if $\dim v = \dim u$ and there exists a \linebreak
$C^*$-homomorphism
$\theta_{HG}: A\longrightarrow B$ such that
$\theta_{HG}(u_{ij})=v_{ij}$, $i,j=1,2,\ldots,\dim u$.\\\\
\end{em}

Notice that $\theta_{HG}$ must be a $C^*$-epimorphism.

Let $H\subset G$ be two compact groups of matrices. The conditions of
Def.1.3 are then satisfied by $\theta_{HG}=i^*$ where $i:H\lo G$ is the
natural embedding. Conversely, let $G$ be a compact group of matrices.
Then each its subgroup in the sense of Def.1.3 is also a
compact subgroup in the
usual sense (up to a
$C^*$-isomorphism).

According to Def.1.3, $S_qU(N), q\in(0,1]$ (see [W4]) is
a subgroup of $S_qU(N+1)$
(we use the identification of Note after Prop.1.2 for the
representation $w=u\sm u^0$ of $S_qU(N)$, cf eq.(1.7) of [NYM]).

The second main notion of the paper is introduced as follows.\\\\

{\sc Definition 1.4}

\begin{em}
Let $X$ be a quantum space and $G$ be a quantum group. We say that a
$C^*$-homomorphism $\Gamma:C(X)\lo C(X)\p C(G)$ is an action of $G$ on $X$ if\\
\noindent a) $(\Gamma\p id)\Gamma=(id\p\Phi_G)\Gamma$\\
\noindent b) $<(I\p y)\Gamma x: x\in C(X), y\in C(G)>=C(X)\p C(G)$
\end{em}\\\\

{\sc Remark 1.} This definition is more restrictive than that used
in [P1]. Nevertheless, Thm.1 and Thm.2 of [P1] remain true if we use instead
Def.1.4 (cf Corollary 1.6).

{\sc Remark 2.} In the classical case (i.e. if $X$ is a usual
compact Hausdorff space and $G$ a
compact group of matrices), Def.1.4 means that $\Gamma = \sigma^*$, where
$\sigma:X\times G\lo X$  is a right continuous action of $G$ on $X$ in the
usual sense (including the condition $\sigma(x,e)=x$ for $x\in X$).

Let $X$ be a quantum space and $G$ be a quantum group. Let us fix a
$C^*$-homomorphism $\Gamma:C(X)\lo C(X)\p C(G)$. We say that a vector
subspace $W\subset C(X)$ corresponds to a representation $v$ of $G$ if
 there exists a basis $e_1,\ldots,e_d$ in $W$ such that
$\dim v=d$ and
$\Gamma e_k=e_m\p v_{mk}$, $k=1,2,\ldots d$. It occurs that if $\Gamma$
is an action of $G$ on $X$ then $C(X)$ can be decomposed into vector
subspaces corresponding to irreducible representations of $G$:\\\\

{\sc Theorem 1.5}

\begin{em}
Let $\Gamma$ be an action of a quantum group $G$ on a quantum space $X$.
We denote $E^{\alpha}=(id\p\rho^{\alpha})\Gamma$,
$W_{\alpha}=E^{\alpha}C(X)\subset C(X)$ for $\alpha\in\hat G$
(see (\ref{1.4})).
Then\\
1) $C(X)=\overline{\sm_{\al\in\hat G} W_{\al}}$\\
2) For each $\al\in\hat G$ there exists a set $I_{\al}$ and vector subspaces
$W_{\al i}$, $i\in I_{\al}$, such that\\
\ \ \ \ a) $W_{\al}=\sm_{i\in I_{\alpha}} W_{\al i}$\\
\ \ \ \ b) $W_{\al i}$ corresponds to $u^{\al}$ for each $i\in I_{\al}$\\
3) Each vector subspace $V\subset C(X)$ corresponding to $u^{\al}$ is
contained in $W_{\al}$\\
4) The cardinal number of $I_{\al}$ doesn't depend on the choice of
$\{W_{\al i}\}_{i\in I_{\al}}$. It is denoted by $c_{\al}$ and called the
multiplicity of $u^{\al}$ in the spectrum of $\Gamma$.\\\\
\end{em}

{\it Proof.}1)2) Set $E^{\al}_{sm}=(id\p\rho^{\al}_{sm})\Gamma:C(X)\lo C(X)$,
$\al\in\hat G$, $s,m=1,2,\ldots,d_{\al}$. Using condition a) of Def.1.4 and
(\ref{1.2}), we get
\begin{equation}
E^{\al}_{sm}E^{\beta}_{ij}
=[id\p(\rho^{\al}_{sm}\p\rho^{\beta}_{ij})\Phi_G]\Gamma=
\delta_{mi}\delta_{\al(\beta)}E^{\beta}_{sj}.\label{1.5}
\end{equation}
By virtue of $<x^{\al}_{sm}:\al\in\hat G, s,m=1,\ldots,d_{\al}>=C(G)$
and condition b) of Def.1.4, we obtain
\begin{equation}
\left.\begin{array}{l}
<E^{\al}_{sm}x:\al\in\hat G, s,m=1,\ldots,d_{\al}, x\in C(X)>\\
=<(id\p h)(I\p y)\Gamma x: y\in C(G), x\in C(X)>=C(X).
\end{array}\right\}\label{1.6}
\end{equation}
Let $W^{\al s}=E^{\al}_{(s)(s)}C(X)$. Using (\ref{1.5}) and (\ref{1.6}),
we get
\begin{equation}
C(X)=\overline{\sm_{\al,s}W^{\al s}}.\label{1.7}
\end{equation}
But $E^{\al}=E^{\al}_{ss}$, hence
\begin{equation}
W_{\al}=\sm^{d_{\al}}_{s=1} W^{\al s},\label{1.8}
\end{equation}
which proves 1). Let $\{e_{\al i1}\}_{i\in I_{\al}}$ be a basis of $W^{\al1}$.
We set $e_{\al is}=E^{(\al)}_{s1}e_{\al i1}$, $s=1,\ldots,d_{\al}$,
$i\in I_{\al}$. In virtue of (\ref{1.5}), $\{e_{\al is}\}_{i\in I_{\al}}$ is a
basis of $W^{\al s}$. Putting $W_{\al i}=
\ \mbox{span}\ \{e_{\al is}: s=1,\ldots,d_{\al}\}$ and using (\ref{1.8}),
we get 2a). Using condition a) of Def.1.4 and (\ref{1.3}), we get
\[ \Gamma e_{\al is}=\Gamma E^{\al}_{(s)(s)}e_{\al is}
=\Gamma(id\p\rho^{\al}_{(s)(s)})\Gamma e_{\al is} \]
\[ =(id\p(id\p\rho^{\al}_{(s)(s)})\Phi_G)\Gamma e_{\al is} \]
\[ =[id\p P^{\al}_{(s)(s)}]\Gamma e_{\al is}\in
C(X)\p\ \mbox{span}\ \{u^{\al}_{js}: j=1,\ldots,d_{\al}\}. \]
Therefore $\Gamma e_{\al is}=x_{(\al)i(s)j}\p u^{\al}_{js}$ for some
$x_{\al isj}\in C(G)$, $j=1,2,\ldots,d_{\al}$. Acting on both sides by
$(id\p\rho^{(\al)}_{k(s)})$, $k=1,2,\ldots,d_{\al}$, we obtain
$E^{(\al)}_{k(s)}e_{\al is}=x_{\al isk}$, hence
$x_{\al isk}=e_{\al ik}$, $\Gamma e_{\al is}=e_{(\al)ij}\p u^{\al}_{js}$,
$s=1,2,\ldots,d_{\al}$, $i\in I_{\al}$. It proves b).

3)4) Let $e_1,e_2,\ldots,e_{d_{\al}}$ form a basis of $V\subset C(X)$
such that $\Gamma e_s=e_j\p u^{\al}_{js}$, $s=1,\ldots,d_{\al}$.
We get $E^{\al}_{r1}e_1=e_r$, $r=1,\ldots,d_{\al}$. Thus
$e_r=E^{\al}_{(r)(r)}e_r\in W^{\al r}\subset W_{\al}$, which proves 3).
Moreover, we see that each decomposition of type 2a) can be obtained as in
proof of 2). Therefore the cardinal number of $I_{\al}$ is equal to
$\dim W^{\al1}$ for each choice of $\{W_{\al i}\}_{i\in I_{\al}}$.
\hfill$\Box$.\\\\

{\sc Corollary 1.6}

\begin{em}
Let $X$ be a quantum space, $G$ be a quantum group and
$\Gamma:C(X)\lo C(X)\p C(G)$ be a $C^*$-homomorphism. Then $\Gamma$ is an
action of $G$ on $X$ iff there exist sets $J_{\al}$, $\al\in\hat G$, and
linearly independent and linearly dense elements $e_{\al mj}$, $\al\in\hat G$,
$m\in J_{\al}$, $j=1,\ldots,d_{\al}$, in $C(X)$ such that
$\Gamma e_{\al mj}=e_{\al ms}\p u^{(\al)}_{sj}$. In that case
$\#J_{\al}=c_{\al}$
if one of these values is finite.\\\\
\end{em}

{\it Proof.\/} `$\Rightarrow$'is contained in Theorem 1.5. Conversely, let
such elements $e_{\al mj}$ be given. Then condition a) of Def.1.4 is satisfied
(it suffices to check it on $e_{\al mj}$), while the condition b) follows from
$e_{\al mk}\p w=\linebreak
 \{I\p[w{(u^{\al T})^{-1}}_{kj}]\}\Gamma e_{(\al)mj}$, where
$w\in C(G)$, $\al\in\hat G$, $m\in J_{\al}$, $k=1,2,\ldots,d_{\al}$.
Moreover (see the proof
of Th.1.5), $W^{\al s}=<e_{\al ms}: m\in J_{\al}>$,
$c_{\al}=\linebreak
\dim<e_{\al m1}: m\in J_{\al}>$, which proves the last statement.
\hfill$\Box$.\\\\

Now we shall find the numbers $c_{\al}$ for the quotient spaces. Let $H$ be
a subgroup of a quantum group $G$. The quotient space $H\backslash G$
is defined by
\[ C(H\backslash G)=\{x\in C(G): (\th_{HG}\p id)\Phi_Gx=I\p x\} \]
(cf [P1, Sec.6]). Similarly as in [P1,Sec.6] we get that
$E_{H\backslash G}=\linebreak
(h_H\p id)(\th_{HG}\p id)\Phi_G$ is a completely bounded
projection from $C(G)$ onto $C(H\backslash G)$. Moreover,
$(E_{H\backslash G}\p id)\Phi_G=\Phi_GE_{H\backslash G}$. Thus we can define
\[ \Gamma_{H\backslash G}=\Phi_{G\mid_{C(H\backslash G)}}:C(H\backslash G)\lo
C(H\backslash G)\p C(G). \]

Let $\al\in\hat G$. The representation $\th_{HG}(u^{\al})$ of the group $H$ can
be decomposed into a direct sum of irreducible representations among which the
trivial one appears with a multiplicity which we denote by $n_{\al}$. Taking a
suitable form of $u^{\al}$ we get
\[ h_H(\th_{HG}(u^{\al}_{ij}))=\left\{
\begin{array}{ll}
\delta_{ij} & \ \mbox{for}\ i=j,\ i=1,2,\ldots,n_{\al},\\
0           & \ \mbox{otherwise}\ .
\end{array}\right. \]
Therefore $E_{H\backslash G}u^{\al}_{mj}=u^{\al}_{mj}$
for $1\leq m\leq n_{\al}$
and $E_{H\backslash G}u^{\al}_{mj}=0$ for $n_{\al}<m\leq d_{\al}$, $j=1,\ldots,
d_{\al}$, $\al\in\hat G$. Hence, $e_{\al mj}=u^{\al}_{mj}$,
$m=1,\ldots,n_{\al}$, $j=1,\ldots,d_{\al}$, $\al\in\hat G$, have the same
properties as in Corollary 1.6 with $J_{\al}=\{1,\ldots,n_{\al}\}$. We
obtain\\\\

{\sc Theorem 1.7}

\begin{em}
Let $H$ be a subgroup of a quantum group $G$. Then
$\Gamma_{H\backslash G}=\Phi_{G\mid_{C(H\backslash G)}}:C(H\backslash G)\lo
C(H\backslash G)\p C(G)$ is an action of $G$ on $H\backslash G$. Moreover,
$c_{\al}=n_{\al}$ (the multiplicity of the trivial representation $(I)$ in the
decomposition of $\th_{HG}(u^{\al})$ into irreducible components).\\\\
\end{em}

{\sc Definition 1.8}

\begin{em}
All the pairs $(H\backslash G,\Gamma_{H\backslash G})$
obtained in the above way (and
the pairs isomorphic to them) are called quotient. Let $\Gamma$ be an action
of a quantum group $G$ on a quantum space $X$. We say that a pair $(X,\Gamma)$
is embeddable if $C(X)\neq\{0\}$ and there exists a faithful
$C^*$-homomorphism $\psi:C(X)\lo C(G)$ such that
$\Phi_G\psi=(\psi\p id)\Gamma$ (cf [VS2]). We say
that $(X,\Gamma)$ is homogeneous if
$c_0=1$.\\\\
\end{em}

{\sc Remark 3.} In the classical case $C(H\backslash G)$ is the commutative
$C^*$-algebra of functions which are constant on the orbits $Hg$ ($g\in G$) of
the subgroup $H$ of $G$. Let $\pi$ be the continuous projection
$\pi:G\lo H\backslash G$. Then $\pi^*$ identifies that $C^*$-algebra with the
$C^*$-algebra of continuous functions on the usual quotient space
$H\backslash G$.
Then $\Gamma$ is identified with $\sigma^*$, where $\sigma$ is the usual right
continuous action of $G$ on $H\backslash G$.

{\sc Remark 4.} In the classical case $(X,\Gamma)$ is homogeneous iff
$X\neq\emptyset$ is homogeneous w.r.t. the action $\sigma$ of the group $G$
(see Remark 2).

{\it Proof.\/} Let $x\in C(X)$. Then $x\in W_0$ iff $\Gamma x=x\p I$ iff
$x(pg)=x(p)$, $p\in X$, $g\in G$ iff $x$ is constant on the orbits $pG$
of $G$.

$\Leftarrow$: If $X\neq\{0\}$ is homogeneous then $pG=G$, $W_0=\ce I$, $c_0=1$.

$\Rightarrow$: $X\neq\{0\}$ since $c_0=1>0$. Assume ad absurdum that $X$ is
not homogeneous. Then there exist $p,p'\in X$ such that $p'\notin pG$. By
the Urysohn lemma there exists $f\in C(X)$ such that $0\leq f\leq 1$,
$f_{\mid_{pG}}=0$, $f(p')=1$. Let $k=(id\p h)\Gamma f$. Then
$k=E^0 f\in W^0=\ce I$. But $k(p)=\int_G f(pg)dg=0$,
$k(p')=\int_G f(p'g)dg>0$. This contradiction proves the homogenity of $X$.
\hfill$\Box$.

A relation among the above notions is given by\\\\

{\sc Proposition 1.9}

Let $\Gamma$ be an action of a quantum group $G$ on a quantum space $X$. Then\\
\noi a) $(X,\Gamma)$ is quotient $\Rightarrow$ $(X,\Gamma)$ is embeddable
$\Rightarrow$ $(X,\Gamma)$ is homogeneous\\
\noi b) In the classical case
$(X,\Gamma)$ is quotient $\lon$ $(X,\Gamma)$ is embeddable $\lon$ $(X,\Gamma)$
is homogeneous\\\\

{\it Proof:\/} a) The first implication holds for $\psi=id:C(H\backslash G)\lo
C(G)$. Let now $(X,\Gamma)$ be embeddable, $x\in C(X)$, $\Gamma x=x\p I$. Then
$\Phi_G\psi(x)=\psi(x)\p I$. Acting on both sides by $id\p h_G$ we get
$\psi(x)=h_G(\psi(x))I\in\ce I$, $x\in\ce I$. Thus $W_0=\ce I$, $c_0=1$.

\noi b) In this case each homogeneous space is (up to a homeomorphism)
quotient, which proves the implications opposite to that of a).
\hfill$\Box$.\\\\

{\sc Remark 5.} Examples of non-compact quantum homogeneous spaces are
given by [W8].

\section{Subgroups and quotient spaces of quantum $SU(2)$ groups.}

\ \ \ \
In this Section we classify the subgroups of quantum groups $SU_q(2)$,
$q\in[-1,1]\setminus\{0\}$. The corresponding quotient spaces are
described (for $q\in(-1,1)\setminus\{0\}$).

First, let us recall that compact subgroups of $SO(3)$ are given by\\
\noi a) $SO(3)$,\\
\noi b) $SO(2)_{{\bf n}}$ (all rotations around the axis given by ${\bf n}$),\\
\noi c) $DO(2)_{{\bf n}}$ (the elements of $SO(2)_{{\bf n}}$ and all rotations
through angle $\pi$ around axes perpendicular to ${\bf n}$),\\
\noi d) $C_{m,{\bf n}}$ (rotations through angles $\frac{2\pi}{m}k$,
$k=0,1,\ldots,m-1$, around axis given by ${\bf n}$), $m=1,2,\ldots$,\\
\noi e) $D_{m,{\bf n},\phi}$ (the elements of $C_{m,{\bf n}}$ and rotations
through angle $\pi$ around $m$ axes in plane $\sigma_{{\bf n}}$
perpendicular to ${\bf n}$, with equal angles between neighbouring axes, where
$\phi$ denotes the angle in $\sigma_{{\bf n}}$ between the projection of
${\bf e}_3$ on $\sigma_{{\bf n}}$ (we take ${\bf e}_1$ instead of ${\bf e}_3$
 if
${\bf n}=\pm{\bf e}_3$) and first axis in anti--clockwise direction),
$m=2,3,\ldots$, $0\leq\phi<\frac{\pi}{m}$,\\
\noi f) $T_{{\bf n},\phi}$ (the symmetries of regular tetrahedron with one of
vertices in direction of ${\bf n}$, where $\phi$ is now measured towards a
projection of an edge starting in this vertex), $0\leq\phi<\frac{2\pi}{3}$,\\
\noi g) $O_{{\bf n},\phi}$ (the symmetries of regular octahedron with
${\bf n},\phi$ defined as in f)), $0\leq\phi<\frac{\pi}{2}$,\\
\noi h) $I_{{\bf n},\phi}$ (the symmetries of regular icosahedron with
${\bf n},\phi$ defined as in f)), $0\leq\phi<\frac{2\pi}5$,

where ${\bf n}$ is a unit vector. We have two opposite choices of ${\bf n}$ for
any subgroup in b)-f)
(in the case of f) the change of sign of {\bf n} corresponds to the inversion
of the tetrahedron)
and many choices of ${\bf n}$ corresponding to the
vertices
of the solid for any subgroup in f)-h) (thus we have 8 choices
for f); $\phi$ is unique for a given ${\bf n}$,
but can depend on its choice); moreover $C_1$ doesn't depend on ${\bf n}$,
$D_2$ depends only on the set of three perpendicular axes; other subgroups are
distinct.

Let $\beta:SU(2)\lo SO(3)$ be the standard continuous two--folded covering:
$[\beta(g)]w=gwg^{-1}$, where $g\in SU(2)$, $w=x\sigma_x+y\sigma_y
+z\sigma_z\simeq(x,y,z)\in{\bf R}^3$.
The compact subgroups $H\subset SU(2)$ fall into two
classes:

\noi 1)$-I\notin H$. Then $\beta(H)$ can't contain $C_{m,{\bf n}}$ for any even
$m$. We must have $\beta(H)=C_{m,{\bf n}}$ for some odd $m$. Then we have
exactly one such subgroup
$H=({\bf Z}_m)_{{\bf n}}\equiv \{R^{2k}: k=0,1,\ldots,m-1\}$, where $R$
is any generator of cyclic group $\beta^{-1}(C_{m,{\bf n}})$ (the
choice of $R$ is irrelevant, opposite choices
of ${\bf n}$ give
the same subgroup, for $m=1$ ${\bf n}$ is irrelevant, other subgroups are
distinct). \\
\noi 2) $-I\in H$. Then all distinct possibilities are given by
$H=\beta^{-1}(W)$, where $W$ is any compact subgroup of $SO(3)$.

Quantum $SU(2)$ groups [W2] are defined as $SU_q(2)=(A,u)$,
$q\in[-1,1]\setminus\{0\}$, where $A$ is the universal $C^*$-algebra generated
by two elements $\alpha,\gamma$ satisfying
\begin{equation}
\left.\begin{array}{ll}
\al^*\al+\gamma^*\gamma=I,  \al\al^*+q^2\gamma^*\gamma=I,
\gamma^*\gamma=\gamma\gamma^*, \\
\alpha\gamma=q\gamma\alpha,  \alpha\gamma^*=q\gamma^*\alpha
\end{array}\right\}\label{2.1}
\end{equation}
and
\[ u=\left(\begin{array}{ll}
\al, & -q\gamma^*\\
\gamma, & \alpha^*
\end{array}\right). \]
For $q=1$ we get the usual $SU(2)$ group. According to [W2, Sec.5],
 all nonequivalent irreducible
representations of $\su$ can be
chosen as \linebreak
$\{d_{\alpha}\}_{\alpha\in{\bf N}/2}$,
$\dim d_{\alpha}=2\al+1$, $d_{\al}\tp d_{\beta}%
\simeq d_{\mid\alpha-\beta\mid}\sm d_{\mid\al-\beta\mid+1}\sm\ldots\sm
d_{\alpha+\beta}$. We can put $d_{1/2}=u$. The classification of subgroups
of $SU_q(2)$, $q\in(-1,1)\setminus\{0\}$, is given by\\\\

{\sc Theorem 2.1}

$\su$, $q\in[-1,1]\setminus\{0\}$, has the following subgroups:

\noi a) $\su=(A,u)$\\
\noi b) $U(1)=(C(S^1),\left[\begin{array}{ll} z,&0\\0,&\bar z\end{array}\right]
)$
where $S^1=\{e^{i\phi}:\phi\in{\bf R}\}$ and $z\in C(S^1)$ is given by
$z(e^{i\phi})=e^{i\phi}$, $\phi\in{\bf R}$.

\noi c) ${\bf Z}_n=(C(Z_n),\left[
\begin{array}{ll}z_{(n)},&0\\0,&\overline{z_{(n)}}
\end{array}\right])$
where $Z_n=\{e^{2\pi ik/n}:k=0,1,\ldots,n-1\}$ and $z_{(n)}\in C(Z_n)$ is
defined by $z_{(n)}(e^{2\pi ik/n})=e^{2\pi ik/n}$, $n=1,2,\ldots$.

For $q\in(-1,1)\setminus\{0\}$ the above list contains all subgroups of $\su$
(up to $C^*$-isomorphisms, without repetitions). \\\\

{\it Proof:\/} a) is obvious.

b) Compact group of matrices $S^1$ corresponds to the quantum group
$(C(S^1),z)$. By Prop.1.2 (for $w=z\sm \bar z$), $U(1)$ is also a quantum
group. The elements $\tilde\al=z$ and $\tilde\gamma=0$ satisfy (\ref{2.1}),
hence $\th_{U(1)SU_q(2)}$ exists, $U(1)$ is a subgroup of $\su$.

c) can be proved analogously.

We shall prove the last statement. Let $H=(B,v)$ be a subgroup of $\su$,
$q\in(-1,1)\setminus\{0\}$. Then
\[ v=\left[\begin{array}{ll}\tilde\alpha,&-q\tilde\gamma^*\\
\tilde\gamma,&\tilde\al^*\end{array}\right] \]
where $\tilde\alpha, \tilde\gamma$ satisfy(\ref{2.1}). Moreover,
$B=C^*(\tilde\alpha,\tilde\gamma)$. A detailed analysis of relations
(\ref{2.1}) shows (cf [W2], [VS1]) that (up to a $C^*$-isomorphism of the
$C^*$-algebra $B$)\\
\noi 1) $\tilde\alpha=\alpha_0\p I_{C(\Delta)}$, $\tilde\gamma=\gamma_0\p U$,
or

\noi 2) $\tilde\alpha=U$, $\tilde\gamma=0$,

\noi where $\alpha_0, \gamma_0\in B(H_{\infty})$, $H_{\infty}$ is a Hilbert
space with an orthonormal basis $f_0,f_1,\ldots$,
\[ \alpha_0 f_m=(1-q^{2m})^{1/2}f_{m-1},\ \gamma_0 f_m=q^mf_m\ (f_{-1}=0),\
m=0,1,\ldots, \]
$U\in C(\Delta)$ is given by $U(e^{i\phi})=e^{i\phi}$ for $e^{i\phi}\in\Delta$
and $\Delta$ is a nonempty compact subset of $S^1$.

In the case of 1) we define unitary operator $D_c\in B(H_{\infty})$ by
$D_cf_k=e^{ick}f_k$, $k=0,1,2,\ldots$, $c\in{\bf R}$. Using
$\Phi_H\tilde\gamma=\tilde\gamma\p\tilde\alpha+\tilde\alpha^*\p\tilde\gamma$
we get
\[ (D_c^*\p id\p D_c\p id)\Phi_H\tilde\gamma(D_c^*\p id\p D_c\p id)^*=
e^{-ic}\Phi_H\tilde\gamma. \]
Therefore
\begin{equation}
Sp(\Phi_H\tilde\gamma)=e^{-ic}Sp(\Phi_H\tilde\gamma).\label{2.2}
\end{equation}
But $\Phi_H\tilde\alpha$ and $\Phi_H\tilde\gamma\neq0$ also satisfy
(\ref{2.1}), hence they are (up to a \linebreak
$C^*$-isomorphism) such as in 1) for some
$\Delta'$. By virtue of (\ref{2.2}), $\Delta'=e^{-ic}\Delta'$ for all
$c\in{\bf R}$, $\Delta'=S^1$. Since $Sp(\Phi_H\tilde\gamma)\subset
Sp\tilde\gamma$, $\Delta'\subset\Delta$, $\Delta=S^1$.
We can identify $H$ with $\su$.

In the case of 2) $\Phi_H\tilde\alpha=\tilde\alpha\p\tilde\alpha$. Therefore
\[
\Delta\cdot\Delta=Sp(\tilde\alpha\p\tilde\alpha)=Sp\Phi_H\tilde\alpha
\subset Sp\tilde\alpha=\Delta.
\]
We get $\Delta=S^1$ or $\Delta=Z_n$, $n=1,2,\ldots$. Hence, $H$ is such as in
b) or c).

These subgroups are distinct since the corresponding $C^*$-algebras are
nonisomorphic. \hfill$\Box$.

{\sc Remark 1.} In the case of $q=1$ $\su=SU(2)=\beta^{-1}(SO(3))$,
\[ U(1)=\left\{\left[
\begin{array}{ll}
e^{i\phi},&0\\0,&e^{-i\phi}
\end{array}\right]:\phi\in{\bf R}\right\}=
\beta^{-1}(SO(2)_{{\bf e}_3}), \]
\[ {\bf Z}_n=\left\{\left[
\begin{array}{ll}
e^{2\pi ik/n}, & 0\\0, & e^{-2\pi ik/n}
\end{array}
\right]:k=0,1,\ldots,n-1\right\} \]
which is $({\bf Z}_n)_{{\bf e}_3}$ for odd $n$
and $\beta^{-1}(C_{n/2,{\bf e}_3})$
for even $n$.

Now we shall classify the subgroups of $G=SU_{-1}(2)$. Some related facts were
already given in [Z]. Here we proceed in a little bit more
complete way. First, analysing the set $Sp(A)$ of unitary
equivalence classes of nondegenerate
irreducible representations
of the $C^*$-algebra $A$ (cf [W2, Remark after Th.A2.3]) we get

{\sc Proposition 2.2}

Let $q=-1$. There exists the surjection $\tau:SU(2)\lo Sp(A)$ such that\\
\noi 1)
$[\tau(w)](\alpha)=a\sigma_x$,
$[\tau(w)](\gamma)=c\sigma_y$
for $ac\neq0$,\\
\noi 2) $[\tau(w)](\alpha)=a$, $[\tau(w)](\gamma)=0$ for $c=0$,\\
\noi 3) $[\tau(w)](\alpha)=0$, $[\tau(w)](\gamma)=c$ for $a=0$,\\
where
\begin{equation}
 w=\left(\begin{array}{ll}
a, & -\bar c\\c, & \bar a
\end{array}\right)\in SU(2).\label{2.3}
\end{equation}
We denote $\pi_w=\tau(w)$ and $\pi_Z=\sm_{w\in Z}\pi_w$ for any
$Z\subset SU(2)$.\\\\

{\sc Remark 2.}
Let $\pi\in Sp(A)$. If $\dim\pi=1$ then
$\tau^{-1}(\pi)$ consists of $1$ element (case 2) or 3) above). If
$\dim\pi=2$ then $\tau^{-1}(\pi)$ consists of four elements of $SU(2)$ which
give equivalent irreducible representations. These elements can be obtained one
from another by change of sign of $a$ or/and $c$.\\\\

Let $S\subset Sp(A)$. We set
\[ \tilde S=\{\pi\in Sp(A): \nm\pi(x)\nm\leq\nm\sm_{\rho\in S}\rho(x)\nm
\ \mbox{for all}\ x\in A\} \]
(cf [Dix, \S 3.1]). Clearly $S\subset\tilde S$, $\tilde{\tilde S} = \tilde S$.
Denoting $Z_S=\tau^{-1}(S)$ and by $\bar Z$ the closure of a subset
$Z\subset SU(2)$ we have

{\sc Proposition 2.3}

\[ Z_{\tilde S}=\overline{Z_S}. \]\\

{\it Proof.\/} "$\supset$": If $w\in\overline{Z_S}$ then there exists a
sequence $Z_S\ni w_n\lo w$. It is easy to check that
\[ \nm\pi_w(x)\nm\leq\ \mathop{sup}_n\nm\pi_{w_n}(x)\nm\leq\nm\sm_{\rho\in S}
\rho(x)\nm,\qquad x\in A, \]
hence $\pi_w\in\tilde S$, $w\in Z_{\tilde S}$.

"$\subset$": If $w\in Z_{\tilde S}$ then $\pi_w\in\tilde S$,
$\nm\pi_w(x)\nm\leq\sup_{z\in Z_S}\ \nm\pi_z(x)\nm$ for all $x\in A$. Let us
first consider the case where $ac\neq0$ (see (\ref{2.3})). Setting
\[ x=8I-\left[(\al^2-a^2)^*(\al^2-a^2)+(\gamma^2-c^2)^*(\gamma^2-c^2)\right],
\]
we get $\sup_{z\in Z_S}\ \nm\pi_z(x)\nm\leq 8=\nm\pi_w(x)\nm$. Hence, there
exists a sequence $z_n\in Z_S$ such that $\nm\pi_{z_n}(x)\nm\lo 8$. One can
prove that (maybe replacing $z_n$ by $z_n'$ with $\tau(z_n)=\tau(z_n')$)
$z_n\lo w$, $w\in\overline{Z_S}$. The other cases can be dealt with in the same
manner (for $c=0$ we set $x=4I-(\al-a)^*(\al-a)-\gamma^*\gamma$, for $a=0$ we
set $x=4I-(\gamma-c)^*(\gamma-c)-\al^*\al$).\hfill$\Box$.\\\\

Let us introduce a new (non--associative) product $*$ in $SU(2)$. We set
$x*y=x\cdot y$ for all $x,y\in SU(2)$ except of the case
\[ x=\left(\begin{array}{ll} 0,&-\bar c\\c,&0\end{array}\right),\
y=\left(\begin{array}{ll} 0,&-\bar c'\\c',&0\end{array}\right),
\mid c\mid=\mid c'\mid=1, \]
when $x*y=-x\cdot y$. We say that $Z\subset SU(2)$ is $\tau-conformable$ if
$Z=Z_S$ for some $S\subset Sp\ A$. The subsets and subgroups of $S_{-1}U(2)$
are characterized by \\\\

{\sc Proposition 2.4} (cf Section 2.3 of [Z])

1. Let $Z$ be a compact, $\tau$-conformable subset of $SU(2)$. Then
$\pi_Z:A\lo \pi_Z(A)$ is a $C^*$-epimorphism. In that way we obtain all
$C^*$-epimorphisms from $A$ (up to $C^*$-isomorphisms of the image, without
repetitions).

2. Let $Z$ be a compact, $\tau$-conformable subset of $SU(2)$ such that
$Z*Z\subset Z$. Then $G_Z=(\pi_Z(A),(\pi_Z(u_{ij}))_{i,j=1}^2)$ is a subgroup
of $S_{-1}U(2)$. In that way we obtain all subgroups of $S_{-1}U(2)$ (up to
$C^*$-isomorphisms, without repetitions).\\\\

{\it Proof.\/} 1. Each $C^*$-epimorphism from $A$ has (up to a
$C^*$-isomorphism of the image) the form $\sm_{\rho\in S} \rho$ for some
$S\subset Sp(A)$ ($A$ is separable).
It is clear that $\nm\sm_{\rho\in S} \rho(a)\nm
=\nm\sm_{\rho\in\tilde S} \rho(a)\nm$ for any $S\subset Sp(A)$ and $a\in A$.
Moreover, if $\nm\sm_{\rho\in S} \rho(a)\nm=\nm\sm_{\rho\in T} \rho(a)\nm$ for
some $S=\tilde S$, $T=\tilde T$ and any $a\in A$ then $S=T$ (since
$S\subset\tilde T$, $T\subset\tilde S$). Passing to the subsets of $SU(2)$ and
using Prop.2.3, we get our statement.

2. Let $\rho',\rho''\in Sp(A)$. It is easy to check that
$(\rho'\p\rho'')\Phi_G$ is unitarily equivalent to $\sm_{\rho\in S}\rho$ where
$Z_S=Z_{\{\rho'\}}*Z_{\{\rho''\}}$. Hence (for $Z$ as in 1.),
$Z*Z\subset Z$ iff $\nm(\pi_Z\p\pi_Z)\Phi_G(a)\nm\leq\nm\pi_Z(a)\nm$ (for all
$a\in A$) iff there exists a $C^*$-homomorphism
$\Psi:\pi_Z(A)\lo\pi_Z(A)\p\pi_Z(A)$ such that
$(\pi_Z\p\pi_Z)\Phi_G=\Psi\pi_Z$ iff $G_Z$ is a subgroup of $S_{-1}U(2)$
(the conditions 1,3 of Def.1.1 for $G$ imply the same properties for $G_Z$).
\hfill$\Box$.\\\\

In the following we assume that $Z$ is a compact, $\tau$-conformable subset of
$SU(2)$ such that $Z*Z\subset Z$. We shall find all such subsets. Let us set
\[ L=\{\left(\begin{array}{ll}
a,&-\bar c\\c,&\bar a
\end{array}\right)\in SU(2):\ ac=0\}. \]
We have two cases.

I. $Z\subset L$. For $\phi\in{\bf R}$ we put
\[ O_{\phi}=\left(\begin{array}{ll} e^{i\phi},&0\\0,&e^{-i\phi}
\end{array}\right),
K_{\phi}=\left(\begin{array}{ll} 0,&-e^{-i\phi}\\e^{i\phi},&0
\end{array}\right). \]
Using the definition of $*$ we find that $Z$ has any of the following distinct
forms:\\
\noi a) $Z=L=\beta^{-1}(DO(2)_{{\bf e}_3})$\\
\noi b) $Z=\{O_{\phi}:\phi\in{\bf R}\}=\beta^{-1}(SO(2)_{{\bf e}_3})$
(then $G_Z\approx U(1)$)\\
\noi c) $Z=\{O_{2\pi k/n}, K_{2\pi k/n+\phi_0-\pi/2}:\ k=0,1,\ldots,n-1\}$
 where $n=1,2,\ldots$,
$0\leq\phi_0<\frac{2\pi}{n}$ ($Z=\beta^{-1}(D_{n/2,{\bf e}_3,\phi_0})$ for
$n$ even, $Z$ is not a group for $n$ odd)\\
\noi d) $Z=\{O_{2\pi k/n}:\ k=0,1,\ldots n-1\}$ where $n=1,2,\ldots$
(then $G_Z\approx {\bf Z}_n$;
$Z=\beta^{-1}(C_{n/2,{\bf e}_3})$ for $n$ even, $Z=({\bf Z}_n)_{{\bf e}_3}$
 for $n$ odd)

II. $Z\nsubset L$. Then $-I\in Z$. Indeed, let $w\in Z$ be such that $ac\neq0$.
Hence, $w^k=w*\cdots*w$ ($k$ times) belongs to $Z$ for any $k=1,2,\ldots$.
Therefore $w^{-1}\in Z$ (cf the argument in [W3, proof of Th.1.5]). But
($\tau$-conformability) $-w\in Z$. We get $-I=-w*w^{-1}\in Z$. Consequently,
the considered set $Z$ must have a form $Z=\beta^{-1}(W)$, where $W$ is any
compact subgroup of $SO(3)$ such that $g_0Wg_0^{-1}\subset W$ (where $g_0$ is
the rotation through angle $\pi$ around the axis $x_3$) and
$W\nsubset DO(2)_{{\bf e}_3}$ (omitting the last condition is equivalent with
adding the cases Ia)-Ib) and the cases Ic)-Id) with even $n$).

We get the following $W$:\\
\noi a) $SO(3)$ (then $G_Z\approx
SU_{-1}(2)$)\\
\noi b) $SO(2)_{{\bf n}}$ with ${\bf n}$ perpendicular to ${\bf e}_3$\\
\noi c) $DO(2)_{{\bf n}}$ with ${\bf n}$ perpendicular to ${\bf e}_3$\\
\noi d) $C_{m,{\bf n}}$ with ${\bf n}$ perpendicular to ${\bf e}_3$,
$m=3,4,\ldots$\\
\noi e) $D_{m,{\bf n},0}$ with ${\bf n}$ perpendicular to ${\bf e}_3$,
$m=3,4,\ldots$\\
\noi e') $D_{m,{\bf n},\pi/2m}$ with ${\bf n}$ perpendicular to ${\bf e}_3$,
$m=2,3,\ldots$\\
\noi f) $T_{\phi}$ (the group of symmetries of regular tetrahedron freely
hanging on a horizontal edge, $\phi$ gives the angle between ${\bf e}_1$ and
this edge in anti--clockwise direction), $0\leq\phi<\pi/2$\\
\noi f)' $T'_{\phi}$ (the group of symmetries of a regular tetrahedron with
one edge in vertical position, the opposite one in horizontal position,
$\phi$ gives the angle between ${\bf e}_1$ and this edge in anti-clockwise
direction), $0\leq\phi<\pi$\\
\noi g) $O_{\phi}$ (the group of symmetries of regular octahedron defined as in
f)), $0\leq\phi<\pi$\\
\noi g') $O_{{\bf n},\phi}$ with ${\bf n}={\bf e}_3$, $0\leq\phi<\pi/2$\\
\noi h) $I_{\phi}$ (the group of symmetries of regular icosahedron defined as
in f)), $0\leq\phi<\pi$

Change of sign of {\bf n} doesn't change $W$ in any of the cases b)-e').
Besides, we have distinct $W$ and distinct $Z=\beta^{-1}(W)$. Combining the
above results with Prop.2.4, we get all non--identical subgroups of
$S_{-1}U(2)$.

{\sc Remark 3.} Let $Z$ be $\tau$-conformable subset of $SU(2)$. Then (using an
argument similar to that in [W3, proof of Lemma 4.8]) we get
$\dim\ \pi_Z(A)=\#Z$ (whether $\#Z\in{\bf N}$ or $\#Z=\infty$). In particular,
under assumptions of Prop.2.4.2, $\dim C(G_Z)=\#Z$. Therefore the above
classification gives us many examples of finite--dimensional $*$-Hopf algebras.

Let us consider the quotient spaces $H\backslash G$ w.r.t. the above
subgroups $H$ of $G=\su$. Our aim is to compute the multiplicity $c_{\al}$ of
$d_{\al}$ in the spectrum of the action of $G$ on $H\backslash G$,
$\al\in{\bf N}/2$. According to Thm 1.7, $c_{\alpha}$ are equal to the
multiplicities $n_{\al}$ of the trivial representation in the decomposition of
$\th_{HG}(d_{\al})$ into irreducible components. We start with\\\\

{\sc Proposition 2.5}

Let $G=\su$, $q\in[-1,1]\setminus\{0\}$, and $H$ be a subgroup of $G$ described
in Thm 2.1. Consider the quotient space $H\backslash G$. Then $c_{\al}$ are
equal\\
a) $c_0=1$, $c_{\al}=0$ for $\al>0$ in the case of $H=\su$\\
b) $c_k=1$, $c_{k+1/2}=0$, $k=0,1,2,\ldots$, in the case of $H=U(1)$\\
c) $c_k=2E(\frac{2k}{m})+1$, $c_{k+1/2}=0$, $k=0,1,2,\ldots$, in the case of
$H={\bf Z}_m$, $m$ even;\\
\ \ \ $c_k=2E(\frac{k}{m})+1$, $c_{k+1/2}=2E(\frac{2k+1}{m})-2E(\frac{k}{m})$,
 $k=0,1,2,\ldots$,\\
\ \ \ in the case of $H={\bf Z}_m$, $m$-odd.

{\it Proof:\/} a) $\th_{HG}(d_{\al})=d_{\al}$, $\al\in{\bf N}/2$, which gives
the trivial representation only for $\al=0$ with $c_0=1$.\\
b) $\th_{HG}(d_{\alpha})\simeq z^{-2\al}\sm z^{-2\al+2}\sm\ldots\sm z^{2\al}$
(see [P1, eq.10]). Thus the trivial subrepresentation occurs only for
$\al\in{\bf N}$ (with multiplicity 1).\\
c) Similarly as in b), $\th_{HG}(d_{\al})\simeq
z^{-2\al}_{(m)}\sm z^{-2\al+2}_{(m)}\sm\ldots\sm z^{2\al}_{(m)}$, hence
$n_{\al}$ is the number of elements in the set $\{-2\al,-2\al+2,\ldots,2\al\}$,
which are divisible by $m$. The precise computation gives the results as in the
formulation of the proposition.\hfill$\Box$.\\

The remaining results concerning $q=\pm1$ will be given in Section 3.

In the following we shall illustrate Prop.1.9 using the results of [P1].
Let $q\in[-1,1]\setminus\{0\}$. We say that $(X,\Gamma)$ is a quantum sphere if
$X$ is a quantum space, $\Gamma$ is an action of $\su$ on $X$ and\\
\noi 1) $c_k=1$, $c_{k+1/2}=0$, $k=0,1,2,\ldots$,\\
\noi 2) $W_1$ generates $C(X)$ (as $C^*$-algebra with unity).

{\sc Note.} The notion of the quantum sphere, which is used in the present
paper is more restrictive (for $q\in(-1,1)\setminus\{0\}$)
 than that in [P1]. The present notion coincides with
the assumption (for general $q\in[-1,1]\setminus\{0\}$) and the condition (i')
 of [P1,Th.2].

In the remaining part of the Section $q\in(-1,1)\setminus\{0\}$. According to
[P1, Th.2], $(X,\Gamma)$ is a quantum sphere iff
$(X,\Gamma)$ is isomorphic to $(S^2_{qc},\sigma_{qc})$ for $c\in[0,\infty]$.
Moreover, the constant $c$ is unique.

In [P1] we also considered $(S^2_{qc(n)},\sigma_{qc(n)})$, where
$c(n)=-q^{2n}/(1+q^{2n})^2$, $n=1,2,\ldots$. They satisfy the definition of the
quantum sphere with 1) replaced by\\
1)${}_n$ $c_k=1$ for $k=0,1,\ldots,n-1$, all other $c_k$ vanish (see [P1,
eq.13b and eq.14]).\\\\

{\sc Proposition 2.6} (cf [P1, Sec.6])

Let $q\in(-1,1)\setminus\{0\}$, $c\in\{c(1),c(2),\ldots\}\cup[0,\infty]$.
Then\\
\noi a) $(S^2_{qc},\sigma_{qc})$ is quotient $\lon$ $c=c(1)$ or $c=0$\\
\noi b) $(S^2_{qc},\sigma_{qc})$ is embeddable $\lon$ $c=c(1)$ or
$c\in[0,\infty]$\\
\noi c) $(S^2_{qc},\sigma_{qc})$ is homogeneous

Thus the implications in Prop.1.9.a cannot be replaced by equivalences.\\\\

{\it Proof:\/} a) It follows from Thm.2.1, Prop.2.5 and the properties of
quantum spheres

b) Due to [P1, Prop.4.II] (cf also [MNW, Corollary 3.8]),
$C(S^2_{qc(n)})\approx\pi_+(C(S^2_{qc(n)}))=B(\ce^n)$, $n=2,3,\ldots$,
which has no characters. Therefore $(S^2_{qc(n)},\sigma_{qc(n)})$ is not
embeddable (otherwise $e\spr\psi$ would be a character, where $e$ is the counit
of $\su$). For $c=c(1)$ we set $\psi I=I$. For $c\in[0,\infty]$ we set
$\psi(e_i)=s_kd_{1,ki}$, $i=-1,0,1$, where $(s_{-1},s_0,s_1)$ equals
$(c^{1/2},1,c^{1/2})$ for $c\in[0,\infty)$ and $(1,0,1)$ for $c=\infty$ (due to
$\overline{s_{-k}}=s_k$, $a_{lm}s_ls_m=\rho I$, $b_{lm,k}s_ls_m=\lambda s_k$,
$k=-1,0,1$, and [P2, eq.5], $\psi(e_i)$ satisfy [P1,eq.2]). It is easy to check
the equation $\Phi\psi=(\psi\p id)\sigma_{qc}$ on the generators $e_i$. The
faithfullness of $\psi$ follows from an argument similar as in the proof
of [P1,Th.1(i)$\Rightarrow$(ii)].

c) is obvious.\hfill$\Box$.\\\\

\section{Quantum $SO(3)$ groups.}

\ \ \ \ In this Section we describe subgroups and quotient spaces of quantum
groups $SO_q(3)$, $q\in[-1,1]\setminus\{0\}$. We treat as well the quotient
spaces of $\su$, $q=\pm1$.

We take $d_1$ in the form
\[ d_1=(d_{1,ij})_{i,j=-1,0,1}=\left(
\begin{array}{lll}
\al^{*2},&-(q^2+1)\al^*\gamma,&-q\gamma^2\\
\gamma^*\al^*,&I-(q^2+1)\gamma^*\gamma,&\al\gamma\\
-q\gamma^{*2},&-(q^2+1)\gamma^*\al,&\al^2\\
\end{array}\right) \]
(see [P1, Sec.2]). According to Prop.1.2,
$SO_q(3)=$\linebreak
$(C^*(\{d_{1,ij},\ i,j=-1,0,1\}),d_1)$, $q\in[-1,1]\setminus\{0\}$, are
quantum groups (cf [P1, Remark 3 after Th.2]). The set of their all
nonequivalent irreducible representations can be chosen as
$\{d_{\al}\}_{\al\in{\bf N}}$ (they are generated by $d_1$). Therefore
\begin{equation}
C(SO_q(3))=<d_{\al,mn}:\ m,n=-\al,-\al+1,\ldots,\al,\ \al\in{\bf N}>=
C({\bf Z}_2\backslash\su).\label{3.0}
\end{equation}
That is generated by the elements
\begin{equation}
K=\gamma^*\gamma,\ A=\al\gamma,\ C=\al\gamma^*,\ G=\gamma^2,\ L=\al^2.
\label{3.1}
\end{equation}
It is easy to check that
\begin{equation}
\left.\begin{array}{llllll}
L^*L=(I-K)(I-q^{-2}K),\ LL^*=(I-q^2K)(I-q^4K),\\
G^*G=GG^*=K^2,\ A^*A=K-K^2,\ AA^*=q^2K-q^4K^2,\\
C^*C=K-K^2,\ CC^*=q^2K-q^4K^2,\ LK=q^4KL,\\
GK=KG,\ AK=q^2KA,\ CK=q^2KC,\ LG=q^4GL,\\
LA=q^2AL,\ AG=q^2GA,\ CA=AC,\ LG^*=q^4G^*L,\\
A^2=q^{-1}LG,\ A^*L=q^{-1}(I-K)C,\ K^*=K.
\end{array}\right\}\label{3.2}
\end{equation}
\ \\

{\sc Proposition 3.1}\nopagebreak

Let $q\in(-1,1)\setminus\{0\}$. Then $C(SO_q(3))$ is the universal
$C^*$-algebra generated by $K,A,C,G,L$ satisfying (\ref{3.2}).\\\\

{\it Proof:\/} It follows from\\\\

{\sc Lemma 3.2}

Let $q\in(-1,1)\setminus\{0\}$ and $\tilde K,\tilde A,\tilde C,\tilde G,
\tilde L$ be bounded operators in a Hilbert space $H$, which satisfy
(\ref{3.2}). Then there exist bounded
operators $\tilde\al,\tilde\gamma$ in $H$ which
satisfy (\ref{2.1}) and (\ref{3.1}).\\\\

{\it Proof:\/} We analyse the representations of (\ref{3.2}).\hfill$\Box$.\\\\

{\sc Remark 1.} Let $q\in(-1,1)\setminus\{0\}$. Then [Ta] (using the language
of Hopf algebras) denotes
$SO_q(3)$ by $SO_{q^2}(3)$ and gives its relationship
with $O_{q^2}(3)$ of [RTF], [Ta].\\\\

{\sc Proposition 3.3} (cf [Ta], [P1, Remark 3 after Th.2], [Z, Sec.2.3])

Let $q\in[-1,1]\setminus\{0\}$. Then $SO_q(3)$ is similar to $SO_{-q}(3)$:
there exists a $C^*$-isomorphism $\rho_q:C(SO_q(3))\lo C(SO_{-q}(3))$ such
that\\
a) $\rho_q(K)=K,\ \rho_q(A)=iA,\ \rho_q(C)=iC,\ \rho_q(G)=G,\ \rho_q(L)=L$,\\
b) $\rho_q(d_1)=Qd_1Q^{-1}$,
where $Q=\ \mbox{diag}\ (1,-i,-1)$.\\\\
{\it Proof:\/} We define $C^*$-homomorphism
$T_q:C(\su)\lo C(SU_{-q}(2))\p B(\ce^2)$ by $T_q(\al)=\al\p\sigma_x$,
$T_q(\gamma)=\gamma\p\sigma_y$ (cf Sec.1.1 of [Z] in the case of $q=-1$). Then
$T_q:C(SO_q(3))\lo C(SO_{-q}(3))\p\ \mbox{span}\ \{I,\sigma_z\}.$
The eigenvalues $1,-1$ of $\sigma_z$ correspond to $C^*$-homomorphisms
$\rho_q,\tilde\rho_q:C(SO_q(3))\lo C(SO_{-q}(3))$. Using (\ref{3.1}), we get
a) and analogous formula for $\tilde\rho_q$, with $i$ replaced by $-i$.
Therefore $\tilde\rho_{-q}\spr\rho_q=id$, $\rho_q\spr\tilde\rho_{-q}=id$,
$\rho_q$ is a $C^*$-isomorphism. The property b) follows from a).
\hfill$\Box$.\\\\

{\sc Remark 2} (cf [P1, Remark 3 after Th.2]). Compact group of matrices
$SO_1(3)$ is the image of $SU_1(2)=SU(2)$ under $p=d_1:SU(2)\lo SO_1(3)$. Since
$d_1$ and $\beta$ are both three-dimensional irreducible representations of
$SU(2)$, there exists a matrix $M\in GL(3,\ce)$ such that
$p(x)=M\beta(x)M^{-1}$, $x\in SU(2)$. Thus $SO_1(3)=MSO(3)M^{-1}$ is similar to
$SO(3)$. We can take
\[ M=\left[
\begin{array}{lll}
-1,&-i,&0\\
0,&0,&1\\
-1,&i,&0
\end{array}\right]. \]
Due to Prop.3.3, we have also $SO_{-1}(3)=Q^{-1}SO_1(3)Q$ is similar to
$SO_1(3)$.\\\\

We want to describe subgroups and quotient spaces of $SO_q(3)$,
$q\in[-1,1]\setminus\{0\}$. We start with\\\\

{\sc Proposition 3.4}

Let $H$ be a subgroup of $G=\su$, $q\in[-1,1]\setminus\{0\}$, such that
${\bf Z}_2$ is a subgroup of $H$ (see Th.2.1.c). We set $\tilde G=SO_q(3)$,
$\tilde H=\linebreak
(\th_{HG}(C(\tilde G)),(\th_{HG}(d_{1,mn}))_{m,n=-1}^1)$. Then
$\tilde H$ is a subgroup of $\tilde G$, $C(\tilde H)=C({\bf Z}_2\backslash H)$.
Moreover, $C(\tilde H\backslash\tilde G)=C(H\backslash G)\subset C(\tilde G)
\subset C(G)$, $\Gamma_{\tilde H\backslash\tilde G}=\Gamma_{H\backslash G}$ and
consequently $c_{\al},\al\in{\bf N}$, are the same for both quotient spaces,
$c_{\al}$, $\al\in{\bf N}+1/2$, vanish.\\\\

{\it Proof:\/} It is easy to check that $\tilde H$ is a quantum group and a
subgroup of $\tilde G$ with $\Phi_{\tilde H}=\Phi_{H\mid_{C(\tilde H)}}$,
\begin{equation}
\th_{\tilde H\tilde G}=\th_{HG\mid_{C(\tilde G)}}.\label{3.3}
\end{equation}
Using
\begin{equation}
\th_{{\bf Z}_2H}\th_{HG}=\th_{{\bf Z}_2G},\label{3.4}
\end{equation}
we get
\[
E_{{\bf Z}_2\backslash H}\th_{HG}d_{\al,mn}=\left\{
\begin{array}{ll}
\th_{HG}d_{\al,mn}&\ \mbox{for}\ \al\in{\bf N}\\
0&\ \mbox{for}\ \al\in{\bf N}+1/2.
\end{array}\right.
\]
Therefore $C({\bf Z}_2\backslash H)=\th_{HG}C(\tilde G)=C(\tilde H)$. By virtue
of (\ref{3.4}), one can obtain
$C(H\backslash G)\subset C({\bf Z}_2\backslash G)=C(\tilde G)$. Therefore, due
to ({\ref{3.3}) and
\begin{equation}
\Phi_{\tilde G}=\Phi_{G\mid_{C(\tilde G)}},\label{3.5}
\end{equation}
$C(\tilde H\backslash\tilde G)=C(H\backslash G)$ as $C^*$-subalgebras of
$C(\tilde G)$. Using (\ref{3.5}), we obtain
$\Gamma_{H\backslash G}=\Gamma_{\tilde H\backslash\tilde G}$. Hence, Corollary
1.6 gives that $c_{\al}$ for $H\backslash G$ equals $c_{\al}$ for
$\tilde H\backslash\tilde G$ if $\al\in{\bf N}$ and $0$ if
$\al\in{\bf N}+1/2$.\hfill$\Box$.\\\\

{\sc Remark 3.} Let us identify $C^*$-isomorphic objects. The assumptions
of Prop.3.4
are fulfilled for the
following subgroups $H$ of $\su$:\\
1) $\su$, $U(1)$, ${\bf Z}_{2m}$ ($m\in{\bf N}$) for
$q\in[-1,1]\setminus\{0\}$\\
2) $\beta^{-1}(W)$ where $W$ is any compact subgroup of $SO(3)$ for $q=1$ (then
$\tilde H=p(H)=MWM^{-1}$, i.e. $\tilde H$ is similar to $W$ under $M$,
hence $c_{\alpha}$, $\alpha\in{\bf N}$, for $H\backslash SU(2)$ are the same
as for $W\backslash SO(3)$, $c_{\alpha}$, $\alpha\in{\bf N}+1/2$, vanish)\\
3) $G_{\beta^{-1}(W)}$ where $W$ is any compact subgroup of $SO(3)$ such that
$g_0Wg_0^{-1}\subset W$ for $q=-1$.\\
 They are not fulfilled in the remaining cases:\\
 1)' ${\bf Z}_{2m+1}$ ($m\in\nn$) for $q\in[-1,1]\setminus\{0\}$\\
 2)' $({\bf Z}_{2m+1})_{\bf n}$ ($m\in\nn$)
 for $q=1$ ($({\bf Z}_{2m+1})_{{\bf e}_3}=
 {\bf Z}_{2m+1}$ for $q=1$)\\
 3)' the cases Ic)-Id) with odd $n$ for $q=-1$\\\\

It occurs that Prop.3.4 gives all subgroups of $SO_q(3)$ for
$q\in(-1,1)\backslash\{0\}$:\\\\

{\sc Theorem 3.5}

$SO_q(3)$, $q\in[-1,1]\setminus\{0\}$, has the following subgroups:\\
a) $SO_q(3)=(C(SO_q(3)),d_1)=\tilde{\su}$\\
b) $SO(2)=(C(S^1),\bar z\sm I\sm z)\approx\tilde{U(1)}$\\
c) $C_n=(C(Z_n),\overline{z_{(n)}}\sm I\sm z_{(n)})\approx
\tilde{{\bf Z}_{2n}}, n=1,2,\ldots$\\
For $q\in(-1,1)\setminus\{0\}$ the above list contains all subgroups of
$SO_q(3)$ (up to $C^*$-isomorphisms, without repetitions).\\\\

{\it Proof:\/} We prove the first statement similarly as in Th.2.1 (in order to
prove the second relations we use the fact that $C^*$-homomorphisms
$\psi:C(U(1))\lo C(U(1))$ and $\psi_n:C(Z_n)\lo C(Z_{2n})$ defined by
$\psi(z)=z^2$, $\psi_n(z_{(n)})=z_{(2n)}^2$, $n=1,2,\ldots$,
are faithful). Let now $H=(B,v)$ be a subgroup of $SO_q(3)$, $0<\mid q\mid<1$.
Then
\[ v=\left[\begin{array}{lll}
\tilde L^*,&-q^{-1}(q^2+1)\tilde C^*,&-q\tilde G\\
\tilde A^*,&I-(q^2+1)\tilde K,&\tilde A\\
-q\tilde G^*,&-q^{-1}(q^2+1)\tilde C,& \tilde L
\end{array}\right] \]
where the elements $\tilde K,\tilde A,\tilde C,\tilde G,\tilde L$ generate $B$
and satisfy (\ref{3.2}). Due to Lemma 3.2, there exist
$\tilde\al,\tilde\gamma$, which satisfy (\ref{2.1}) and (\ref{3.1}).
Let us
replace $\tilde\alpha$ by $\tilde\alpha\sm(-\tilde\alpha)$ and $\tilde\gamma$
by $\tilde\gamma\sm(-\tilde\gamma)$
(it changes $H$ only up to a $C^*$-isomorphism).
Hence, $Sp\ \tilde\gamma=-Sp\tilde\gamma$.
Moreover, we can assume that $\tilde\al,\tilde\gamma$ have one of two
forms given in the proof of Th.2.1. In the first case we use the equality
\[ \Phi_H\tilde G=\tilde L^*\p\tilde G+(1+q^{-2})\tilde C^*\p\tilde A+
\tilde G\p\tilde L. \]
Similarly as for $\su$ we get
$Sp\ \Phi_H\tilde G=Sp\ \tilde G=Sp\ G$. Thus $Sp\ \tilde\gamma=Sp\ \gamma$,
$\Delta=S^1$. We can identify $H$ with $SO_q(3)$. In the second case
$\tilde K=\tilde A=\tilde C=\tilde G=0$, while $\tilde L=U^2$. But
$\Phi_H\tilde L=\tilde L\p\tilde L$, hence $Sp\ \tilde L=S^1$ or
$Sp\ \tilde L=Z_n$, $n=1,2,\ldots$. We get (up to a $C^*$-isomorphism)
the cases b) or c). These subgroups are distinct since the corresponding
$C^*$-algebras are nonisomorphic. \hfill$\Box$.\\\\

Combining Prop.2.5 with Prop.3.4, one gets the values of $c_{\alpha}$,
$\alpha\in{\bf N}$, for those subgroups.

Under similarities $SO_{-1}(3)\sim SO_1(3)\sim
SO(3)$, compact subgroups
of $SO_{-1}(3)$,
$SO_1(3)$ and $SO(3)$ are in one to one correspondence,
 hence $c_{\al}$, $\al\in{\bf N}$, are the same for quotient
spaces in all three cases. Those values are given by\\\\

{\sc Proposition 3.6}

Let $W$ be a compact subgroup of $SO(3)$. Consider the quotient space
$W\backslash SO(3)$. Then $c_k$, $k\in{\bf N}$, are equal\\
a) $c_0=1$, $c_k=0$ for $k>0$ in the case of $W=SO(3)$\\
b) $c_k=1$ for $k\in{\bf N}$ in the case of $W=SO(2)_{{\bf n}}$\\
c) $c_k=1$ for $k\in2{\bf N}$, $c_k=0$ for $k\in2{\bf N}+1$ in the case of
$W=DO(2)_{{\bf n}}$\\
d) $c_k=2E(k/m)+1$ in the case of $W=C_{m,{\bf n}}$\\
e) $c_k=E(k/m)+\delta_k$ ($\delta_k=1$ for $k\equiv0$ (mod.2),
$\delta_k=0$ otherwise) in the case of $W=D_{m,{\bf n},\phi}$\\
f) $c_k=E(k/6)+\delta_k$ ($\delta_k=1$ for $k\equiv0,3,4$ (mod.6),
$\delta_k=0$ otherwise) in the case of $T_{{\bf n},\phi}$\\
g) $c_k=E(k/12)+\delta_k$ ($\delta_k=1$ for $k\equiv0,4,6,8,9,10$ (mod.12),
$\delta_k=0$ otherwise) in the case of $O_{{\bf n},\phi}$\\
h) $c_k=E(k/30)+\delta_k$ ($\delta_k=1$ for $k\equiv0, 6, 10, 12, 15, 16,
18, 20, 21, 22, 24, 25, 26,$ $27, 28$ (mod.30), $\delta_k=0$ otherwise)
in the case of $I_{{\bf n},\phi}$.\\\\

{\it Proof:\/} Due to Th.1.7, $c_k$ is equal to the multiplicity of the
trivial representation of $W$ in $\th_{WSO(3)}d_k$. Let
$\chi_k=Tr(\th_{WSO(3)}d_k)$ and $dw$ denotes the Haar measure on $W$.
Then $c_k=(\chi_0\mid\chi_k)=\int_W\chi_k(w)dw$. Using the formula
$\chi_k(w)=\sum_{s=-k,-k+1,\ldots,k} e^{is\phi}$ (where $\phi$ is the
angle of the rotation $w$), we get the above results.\hfill$\Box$.\\\\

Now we pass to the description of quotient spaces for $\su$, $q=\pm1$.
The cases 1), 1)',2) of Remark 3 are already solved. The case 3) is given
by\\\\

{\sc Proposition 3.7}

Let $W$ be a compact subgroup of $SO(3)$ such that $g_0Wg_0^{-1}\subset W$.
Then\\
1.(cf [Z, Sec.2.3]) Under similarities
\[ SO_{-1}(3)\sim^{\rho_1}SO_1(3)\sim SO(3) \]
we have
\[ \tilde{G_{\beta^{-1}(W)}}\sim\tilde{\beta^{-1}(W)}\sim W \]
(up to $C^*$-isomorphisms).\\
2. The quotient spaces $G_{\beta^{-1}(W)}\backslash SU_{-1}(2)$,
$\beta^{-1}(W)\backslash SU(2)$, $\tilde{G_{\beta^{-1}(W)}}\backslash
SO_{-1}(3)$, $\tilde{\beta^{-1}(W)}\backslash SO_1(3)$, $W\backslash
SO(3)$ have the same coefficients $c_{\al}$, $\al\in{\bf N}$ ($c_{\al}$,
$\al\in{\bf N}+1/2$ vanish for the first two quotient spaces).\\\\

{\it Proof:\/} 1. We denote $Z=\beta^{-1}(W)\subset SU(2)$. We shall prove
the first similarity.
One has
\[ \th_{\tilde{G_Z}SO_{-1}(3)}\rho_1(d_1)=\th_{G_ZSU_{-1}(2)}\rho_1
\left[\begin{array}{lll}
L^*,&-2C^*,&-G\\
A^*,&1-2K,&A\\
-G^*,&-2C,&L
\end{array}\right] \]
\[ =\sm_{w\in Z}\pi_w
\left[\begin{array}{lll}
L^*,&2iC^*,&-G\\
-iA^*,&1-2K,&iA\\
-G^*,&-2iC,&L
\end{array}\right]\approx
\sm_{w\in Z}
\left[\begin{array}{lll}
\bar a^2,&2\bar ac\sigma_z,&-c^2\\
-\bar a\bar c\sigma_z,&1-2\bar cc,&-ac\sigma_z\\
-\bar c^2,&2a\bar c\sigma_z,&a^2
\end{array}\right] \]
\[ \approx\sm_{w\in Z}
\left[\begin{array}{lll}
\bar a^2,&-2\bar ac,&-c^2\\
\bar a\bar c,&1-2\bar cc,&ac\\
-\bar c^2,&-2a\bar c,&a^2
\end{array}\right]
\approx\th_{ZSU(2)}d_1=\th_{\tilde ZSO_1(3)}d_1 \]
(we used the fact $\left(\begin{array}{ll}
a,&-\bar c\\
c,&\bar a
\end{array}\right)\in Z\ \Rightarrow\left(\begin{array}{ll}
a,&\bar c\\
-c,&\bar a
\end{array}\right)\in Z$).
Thus $\lambda\th_{\tilde ZSO_1(3)}=\th_{\tilde{G_Z}SO_{-1}(3)}\rho_1$ for some
$C^*$-isomorphism $\lambda:C(\tilde Z)\lo C(\tilde{G_Z})$.
Therefore
$\tilde{G_{\beta^{-1}(W)}}\sim\tilde{\beta^{-1}(W)}$. The second
similarity is given in Remark 3.\\
2. It follows from 1. and Prop. 3.4. \hfill$\Box$.\\\\

In the case 2)' the coefficients $c_{\al}$, $\al\in{\bf N}/2$, for
$({\bf Z}_{2m+1})_{{\bf n}}\backslash SU(2)$ are the same as in Prop.
2.5 in the
case of $H={\bf Z}_{2m+1}$
($({\bf Z}_{2m+1})_{{\bf n}}$ are similar one to another under
unitary matrices; such similarities don't change $SU(2)$).
It remains to consider the case 3)': the case Id) with odd $n$ is covered
by the case 1)', while the case Ic) with odd $n$ is solved by\\\\

{\sc Proposition 3.8}

Let $Z=\{O_{2\pi k/n}, K_{2\pi k/n+\phi_0-\pi/2}: k=0,1,\ldots,n-1\}$ where
$0\leq\phi_0<2\pi/n$, $n$ is odd. Consider the quotient space
$G_Z\backslash SU_{-1}(2)$. Then $c_l=E(l/n)+\delta_l$,
$c_{l+1/2}=E(\frac{2l+1}{n})-E(\frac{l}{n})$ where $\delta_l=1$ for
$l$ even, $\delta_l=0$ for $l$ odd, $l\in{\bf N}$.\\\\

{\it Proof:\/} Since $Z\subset L$, $C(G_Z)$ is commutative. Using Th.1.5
of [W3], we can identify $G_Z$ with $\{O_{2\pi k/n},
S_{2\pi k/n +\phi_0-\pi/2}:
k=0,1,\ldots,n-1\}\subset U(2)$, where
\[ S_{\phi}=\left(\begin{array}{ll}
0,&e^{-i\phi}\\
e^{i\phi},&0
\end{array}\right). \]
Then $c_l=\int_{G_Z}\chi_l(w)dw$, $l\in{\bf N}/2$, where
$\chi_l=Tr(\theta_{G_ZSU_{-1}(2)}d_l)$ and $dw$ denotes the Haar measure on
$G_Z$. Using the formula $\chi_l\chi_{1/2}=\chi_{l-1/2}+\chi_{l+1/2}$,
$l\geq1/2$ (it follows from $d_l\tp d_{1/2}\simeq d_{l-1/2}\sm d_{l+1/2}$),
we get $\chi_l(O_{\phi})=\sum_{s=-2l,-2l+2,\ldots,2l} e^{is\phi}$,
$l\in{\bf N}/2$, $\chi_l(S_{\phi})=(-1)^l$ for $l\in{\bf N}$, $\chi_l(S_{\phi})
=0$ for $l\in{\bf N}+1/2$. An easy computation gives the
result.\hfill$\Box$.\\\\

{\sc Remark 4} (cf [P1, Remark 3 after Th.2]). Let
$(X_{q\la\rho},\sigma_{q\la\rho})$ be as in [P1, Sec.3],
$q\in[-1,1]\setminus\{0\}$, and $e_k$, $k=-1,0,1$, be the corresponding
generators. Then
\[ \sigma_{q\la\rho}:C(X_{q\la\rho})\lo
C(X_{q\la\rho})\p C(SO_q(3))\subset C(X_{q\la\rho})\p C(SU_q(2)). \]
We set
\[ \sigma'_{q\la\rho}=(id\p\rho_q)\sigma_{q\lambda\rho}:C(X_{q\la\rho})\lo
C(X_{q\la\rho})\p C(SO_{-q}(3)) \]
\[ \subset C(X_{q\la\rho})\p C(SU_{-q}(2)). \]
The $C^*$-isomorphism $\Lambda:C(X_{q\la\rho})\lo C(X_{-q\la\rho})$ given
by $\Lambda(e_{-1})=-ie_{-1}$, $\Lambda(e_0)=e_0$, $\Lambda(e_1)=ie_1$
identifies $(X_{q\la\rho},\sigma'_{q\la\rho})$ with
$(X_{-q\la\rho},\sigma_{-q\la\rho})$. Therefore, using [P1, Remark 2 after
Th.2], we get that\\
a) $(S^2_{\pm10},\sigma_{\pm10})=(X_{\pm101},\sigma_{\pm101})$ is a unique
(up to an isomorphism) quantum sphere for $q=\pm1$\\
b) The above object and
$(X_{\pm1,1,(l^2-1)/4},\sigma_{\pm1,1,(l^2-1)/4})$, $l=2,3,\ldots$, are
unique (up to an isomorphism) objects which satisfy the assumptions and
the condition (i) of [P1, Th.1] for $q=\pm1$.

Using Prop.3.4, Th.3.5 and [P1, Sec.6], we get the following realisations for
the quotient quantum sphere (with the corresponding action of the group)
\[ (X_{q,1-q^2,1},\sigma_{q,1-q^2,1})=(S^2_{q0},\sigma_{q0})\approx
U(1)\backslash\su = SO(2)\backslash SO_q(3), \]
$q\in[-1,1]\setminus\{0\}$.\\\\

{\sc Remark 5.} Here we use the terminology of [P5]. Let
$q\in[-1,1]\setminus\{0\}$, $c\in[0,\infty]$
($c=0$ for $q=\pm1$). Then $\Lambda$ of Remark 4
identifies $\ap_c$ for $q$ with $\ap_c$ for $-q$. Using that
identification one can check that $(\sh,\sigma^{\w},d,*)$ is a
$\two$-dimensional exterior algebra on $S^2_{qc}$, invariant w.r.t.
$\sigma_{qc}$ iff $(\sh,(id\p\rho_q)\sigma^{\w},d,*)$ is a
$\two$-dimensional exterior algebra on $S^2_{-qc}$, invariant w.r.t.
$\sigma_{-qc}$. The same holds if we restrict ourselves to
$S^{\w0}\sm\ldots S^{\w k}$ for some $k=1,2,\ldots$,
(with suitable
restrictions of all structures in ${\bf S}^{\w}$, without $*$ or with $*$)
instead of ${\bf S}^{\w}$. That and [P5, Theorem] for $q=1$ prove that [P5,
Theorem] holds also for $q=-1$ (cf the remarks at the beginning of [P5,
Sec.2]).\\\\

{\sc Remark 6.} The results of the paper give a proof that the Haar
measure is faithful for all $\su$, $SO_q(3)$, $q\in[-1,1]\setminus\{0\}$
(the proof in [P1, Sec.4] is valid only for $\su$,
$q\in(-1,1)\setminus\{0\}$, the case of $SU_{-1}(2)$ follows from
[Z, Sec.2.3]). Indeed, let $G=(A,u)$ be a quantum group, $h$ be the Haar
measure, $J=\{b\in A:\ h(b^*b)=0\}$ be the corresponding closed two--sided
ideal, $\pi:A\lo A/J$ be the canonical projection and $u_r=(id\p\pi)u$.
Then, according to [W3, p.656], $G_r=(A/J,u_r)$ is a quantum subgroup of
$G$, $\{\pi(u^{\alpha})\}_{\alpha\in\hat G}$ is
the set of all nonequivalent irreducible unitary representations of
$G_r$. Thus $c_{\al}=\delta_{0\alpha}$
for the quotient space $G_r\backslash G$. Therefore, for $G=\su$ or
$SO_q(3)$, our results show that $G_r=G$ (up to a $C^*$-isomorphism),
$J=\{0\}$, $h$ is faithful.\\\\

{\sc Remark 7.} Throughout the paper we dealt with the right actions. We say
that a $C^*$-homomorphism $\Gamma: C(X)\lo C(G)\p C(X)$ is a left action of
a quantum group $G$ on a quantum space $X$ if\\
a) $(id\p\Gamma)\Gamma=(\Phi_G\p id)\Gamma$\\
b) $<(y\p I)\Gamma x: x\in C(X), y\in C(G)> = C(G)\p C(X).$\\
In that case a vector subspace $W\subset C(X)$ corresponds to a
representation $v$ of $G$ if there exists a basis $e_1,\ldots, e_d$ in $W$
such that $d=\dim v$ and $\Gamma e_k = v_{km}\p e_m$, $k=1,2,\ldots d$. It
occurs that for $G=SU_q(2)$ or $G=SO_q(3)$, $q\in[-1,1]\setminus\{0\}$,
there exists a bijective correspondence between the right and the left actions
on any quantum space $X$: if $\Gamma$ is a right action then
$\Gamma'=(\lambda\p id)\sigma\Gamma$ is a left action ($C^*$-isomorphisms
$\lambda: C(G)\lo C(G)$ and $\sigma:C(X)\p C(G)\lo C(G)\p C(X)$ are given by
$\lambda(\alpha)=\alpha$, $\lambda(\gamma)=\gamma^*$, $\sigma(x\p y)=
y\p x$, $x\in C(X)$, $y\in C(G)$; one has $\Phi_G\lambda=
(\lambda\p\lambda)\sigma\Phi_G$, $e_G\lambda=e_G$). That correspondence gives
a bijective correspondence between right and left quantum spheres (defined
analogously as in the present paper).

There exist matrices $S_l\in M_{2l+1}({\bf C})$, $l=0,1,\ldots$, such that
$\lambda(d_l)^T=(S_l^T)^{-1}d_lS_l^T$. We can take
\[ S\equiv S_1=(S_{1,ij})_{i,j=-1,0,1}=\left(
\begin{array}{lll}
1,& 0,& 0\\
0,& -q(q^2+1)^{-1}, & 0\\
0,& 0,& 1
\end{array}\right). \]
Let us consider the right quantum spheres $(S^2_{qc},\sigma_{qc})$,
$q\in[-1,1]\setminus\{0\}$, $c\in[0,\infty]$ (for $q=\pm1$ we have $c=0$),
and their generators $e_k$, $k=-1,0,1$. We denote the corresponding left
quantum spheres by $({S^2_{qc}}^{\prime},\sigma_{qc}^{\prime})$. Then
${S^2_{qc}}^{\prime}=S^2_{qc}$.
Setting $e_k'=S_{ik}e_i$, one has $\sigma_{qc}^{\prime}e_k'=d_{1,kj}\p e_j'$,
$k=-1,0,1$. We put $a'=(S^{-1}\p S^{-1})a\in M_{9\times 1}({\bf C})$,
$b'=(S^{-1}\p S^{-1})bS$, $c'=(S^{-1}\p S^{-1})cS_2$. Then
\[ a^{\prime T} (b^{\prime T}, c^{\prime T}, \mbox{resp.})
\mbox{ intertwines } d_1\tp d_1 \mbox{ with }
d_0 (d_1, d_2, \mbox{resp.}). \]
One can use $X_{q\lambda\rho}'=X_{q\lambda\rho}$, $\sigma_{q\lambda\rho}'$,
$e_k'$, $a'$, $b'$, $c'$, instead of $X_{q\lambda\rho}$,
$\sigma_{q\lambda\rho}$, $e_k$, $a$, $b$, $c$. Cf also [NM].

All the results of the present paper preceding this Remark, [P1, Thm 1 and
Thm 2], [P2, Thm 1] and [P5, Thm] remain true if we use the left actions
and the left quantum spheres (after appropriate modifications, e.g.
eq. (\ref{1.5}) takes form
$E^{\alpha}_{sm}E^{\beta}_{ij}=
[(\rho^{\beta}_{ij}\p\rho^{\alpha}_{sm})\Phi_G\p id]\Gamma=
\delta_{js}\delta_{\alpha(\beta)}E^{\beta}_{im}$,
$e_{\alpha is}=E^{(\alpha)}_{1s}e_{\alpha i1}$,
$E_{G\slash H}=(id\p h_H)(id\p\theta_{HG})\Phi_G$, for $c\in[0,\infty]$
we use the embedding $\psi'=\lambda\psi$:
$\psi'(e_i')=d_{1,ik}s_k'$ with $s_k'=S_{jk}s_j$, $\psi'$ gives the
isomorphism $({S^2_{q0}}^{\prime},
\sigma_{q0}^{\prime})\approx SU_q(2)\slash U(1)$,
$q\in[-1,1]\setminus\{0\}$, $c=0$).\\\\

{\it Acknowledgements.\/}
I am very indebted to Prof.S.L. Woronowicz for stimulating discussions.
A part of this work was done during my stay in RIMS, Kyoto University.
I am grateful to Prof. H. Araki for his kind hospitality there.
I also thank a referee for his useful remarks.\\\\

\end{document}